\documentclass[prl,twocolumn]{revtex4-1}

\usepackage{graphicx}

\newcommand{\kms}{\ensuremath{\mathrm{km}\,\mathrm{s}^{-1}}}

\newcommand{\accunits}{\ensuremath{\mathrm{m}\,\mathrm{s}^{-2}}}
\newcommand{\MLsun}{\ensuremath{\mathrm{M}_{\odot}/\mathrm{L}_{\odot}}}
\newcommand{\Lsun}{\ensuremath{\mathrm{L}_{\odot}}}
\newcommand{\Msun}{\ensuremath{\mathrm{M}_{\odot}}}

\newcommand{\lumdens}{\ensuremath{\mathrm{L}_{\odot}\,\mathrm{pc}^{-2}}}

\newcommand{\gobs}{\ensuremath{\mathrm{g}_{\mathrm{obs}}}}
\newcommand{\gbar}{\ensuremath{\mathrm{g}_{\mathrm{bar}}}}

\newcommand{\ML}{\ensuremath{\Upsilon_{\star}}}
\newcommand{\azero}{\ensuremath{\mathrm{g}_{\dagger}}}

\begin{document}

\title{The Radial Acceleration Relation in Rotationally Supported Galaxies}

\author{Stacy S. McGaugh} 
\author{Federico Lelli} 
\affiliation{Department of Astronomy, Case Western Reserve University, 10900 Euclid Avenue, Cleveland, OH 44106, USA}
\author{James M. Schombert} 
\affiliation{Department of Physics, University of Oregon, Eugene, OR 97403, USA}

\date{\today}

\begin{abstract}
We report a correlation between the radial acceleration traced by rotation curves and that predicted by the observed distribution of baryons. 
The same relation is followed by 2693 points in 153 galaxies with very different morphologies, masses, sizes, and gas fractions. 
The correlation persists even when dark matter dominates. 
Consequently, the dark matter contribution is fully specified by that of the baryons. 
The {observed} scatter is small and {largely dominated by} observational uncertainties. 
This radial acceleration relation is tantamount to a natural law for rotating galaxies.
\end{abstract}

\pacs{95.30.Sf,  95.35.+d, 98.52.Nr, 98.52.Wz, 04.50.Kd}

\maketitle

\section{Introduction}

The missing mass problem in extragalactic systems is well established.   
The observed gravitational potential cannot be explained by the stars and gas.
A classic example is that the rotation curves of disk galaxies become approximately flat 
($V \approx$ constant) when they should be falling in a Keplerian ($V \propto R^{-1/2}$) fashion \cite{vera,bosma}.  

The flatness of rotation curves is only the beginning of the story of the mass discrepancy in galaxies.
For example, the baryonic mass of a galaxy
(the sum of its stars and gas: $M_{\mathrm{bar}} = M_{\star} + M_g$) correlates with the amplitude
of the flat rotation velocity $V_f$. This baryonic Tully-Fisher relation \cite{TForig,btforig,verhTF}
is a simple scaling relation ($M_{\mathrm{bar}} \propto V_f^4$) with no apparent dependence on other properties
like galaxy size \cite{CR,myPRL} or surface brightness \cite{zwaanTF,MdB98a}. 
It has remarkably little intrinsic scatter \cite{M11,MS15,LMS2016}.  This implies a strong connection
between the baryons and the physics that sets $V_f$.

There are further indications on a connection between baryons and dynamics.
Features like spiral arms have corresponding bumps in rotation curve \cite{renzorule}.
The ratio of dark to baryonic mass is known to depend on acceleration \cite{M04,GalRev}.
Here we demonstrate the existence of a quantitative relation between the acceleration due to
the baryons and that due to the total mass.
A key advance is that near-infrared photometry provides a direct link between starlight and stellar mass:
the relation follows from the data with no adjustable parameters.

\section{Data}

Galaxies come in a wide range of morphologies, masses, sizes, and densities. 
Generically they are either pressure supported (ellipticals) or rotationally supported (spirals and irregulars).
Here we consider rotationally supported systems where the rotation curve provides a direct tracer of the centripetal acceleration:
\begin{equation}
\gobs = \frac{V^2(R)}{R} = \left|\frac{\partial{\Phi_{\mathrm{tot}}}}{\partial{R}}\right|,
\label{eq:gobs}
\end{equation}
where $\Phi_{\mathrm{tot}}$ is the gravitational potential and $V(R)$ is the full, resolved rotation curve.
We do not consider pressure supported elliptical galaxies for which the derivation of the potential is more complex,
but there are indications that they may obey a similar phenomenology \cite{Scarpa2006,ellipticalBTF,Serra2016}.

\subsection{Galaxy Sample}

We employ the new Spitzer Photometry and Accurate Rotation Curves (SPARC) database \cite{SPARCI}.
SPARC is a sample of 175 disk galaxies representing all rotationally supported morphological types.
It includes near-infrared (3.6$\mu$m) observations that trace the distribution of stellar mass and 21 cm observations that trace the atomic gas.
The 21 cm data also provide velocity fields from which the rotation curves are derived. In some cases these are supplemented
by high spatial resolution observations of ionized interstellar gas.
SPARC is the largest galaxy sample to date with spatially resolved data on the distribution of both stars and gas
as well as rotation curves for every galaxy.
See \cite{SPARCI} for a complete description of the sample and associated data.

\begin{figure}
\includegraphics[width=3.5in]{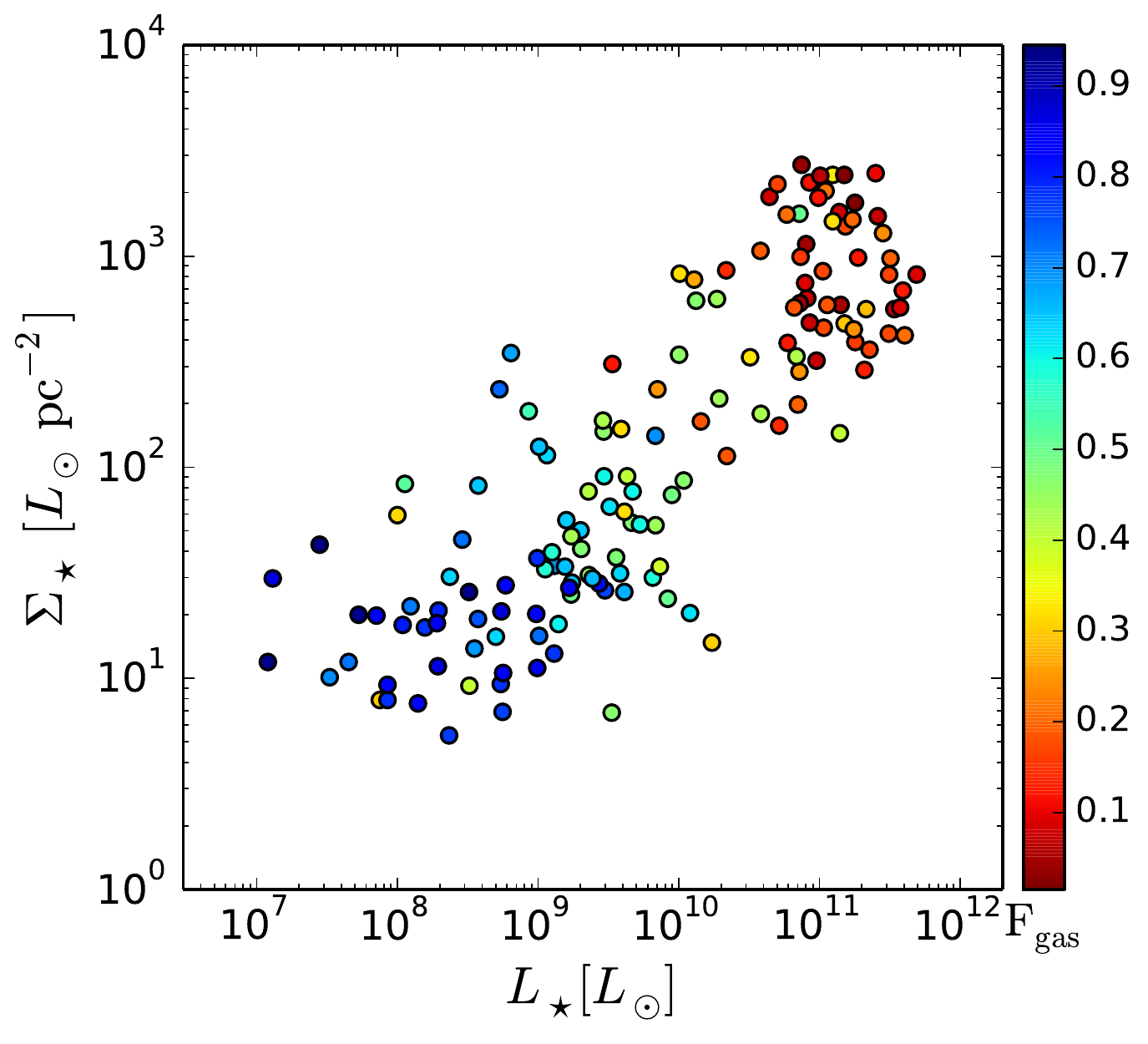}
\caption{The distribution of SPARC galaxies in luminosity and effective surface brightness.
Points are coded by gas fraction (side bar). SPARC samples all known properties of rotationally supported galaxies,
from low to high mass, low to high surface brightness, and negligible to dominant gas content.
\label{SPARCsample}}
\end{figure}

For the purposes of this study, we apply a few modest quality criteria.
Ten face-on galaxies with $i < 30^{\circ}$ are rejected to minimize $\sin(i)$ corrections to the observed velocities.
Twelve galaxies with asymmetric rotation curves that manifestly do not trace the equilibrium gravitational potential are rejected.
This leaves a sample of 153 galaxies. Of the many resolved points along the rotation curves of these galaxies, we require 
a minimum {precision} of 10\% in velocity. 
This retains 2693 data points out of 3149. Dropping this last requirement has no affect
on the result; it merely increases the scatter as expected for less accurate data.

SPARC extends over an exceptional range of physical properties (Fig.~\ref{SPARCsample}).  It includes galaxies 
with rotation velocities $20 < V_f < 300\;\kms$, luminosities $10^7 < L_{[3.6]} < 5 \times 10^{11}\;\Lsun$,
gas masses $10^7 < M_{\mathrm{gas}} < 5 \times 10^{10}\;\Msun$, 
gas fractions $0.01 < F_{\mathrm{gas}} < 0.97$, half-light radii $0.3 < R_{1/2} < 5$ kpc, and
effective surface brightnesses $5 < \Sigma_{\star} < 3 \times 10^3\;\lumdens$. This range extends from some of largest individual 
galaxies known to many of the smallest.  
{SPARC samples well the range of properties of disk galaxies found in complete samples \cite{GAMA,Bradford2015,RESOLVE}.}
Low mass and low surface brightness galaxies are {particularly} well represented in SPARC, 
in contrast to flux selected samples that are typically restricted to $M_{\star} > 10^9\;\Msun$ and $V_f > 100\;\kms$. 

All galaxies have been observed \cite{SMpaper4} at 3.6 $\mu$m with the \textit{Spitzer Space Telescope}. 
This provides the most accurate available tracer of the stellar mass \cite{meidt,SMpaper3,MS14}.
Critically, there is little variation in the conversion from starlight to stellar mass \cite{MS15,DiskMass7}:
what you see in the near-infrared is what you get for the gravitational potential of the stars.
We have uniformly analyzed all of the photometric data using the procedures described in \cite{SMpaper4}.

Galaxies were selected for the availability of resolved 21 cm data. 
These interferrometric data are expensive in both telescope time and labor, and are the limiting factor on sample size.
These rotation curves represent the fruits of decades of work by an entire community of radio astronomers 
(see references in \cite{SPARCI}).  SPARC provides the broadest view of disk galaxies currently available.

\subsection{The Gravitational Potentials of Baryons}

Baryonic mass models are constructed from the observed distribution of stars and gas.
Azimuthally averaged surface brightness profiles are 
converted to surface density assuming a constant mass-to-light ratio for the stars.
The same prescription is used in all galaxies (see below).
The conversion for gas is known from the physics of the spin-flip transition of atomic hydrogen \cite{Draine}.
The {atomic} gas profiles are scaled up by a factor of 1.33 to account for the cosmic abundance of helium \cite{BBNHelium}.
We make the customary assumption that galactic disks have a small but finite thickness to obtain
the 3D density $\rho_{\mathrm{bar}}$ \cite{SPARCI}. While it is important to account for the cylindrical rather
than spherical geometry of disks \cite{Casertano,BT}, the results are not sensitive to the detailed implementation of disk thickness.

\begin{figure*}
\includegraphics[width=7in]{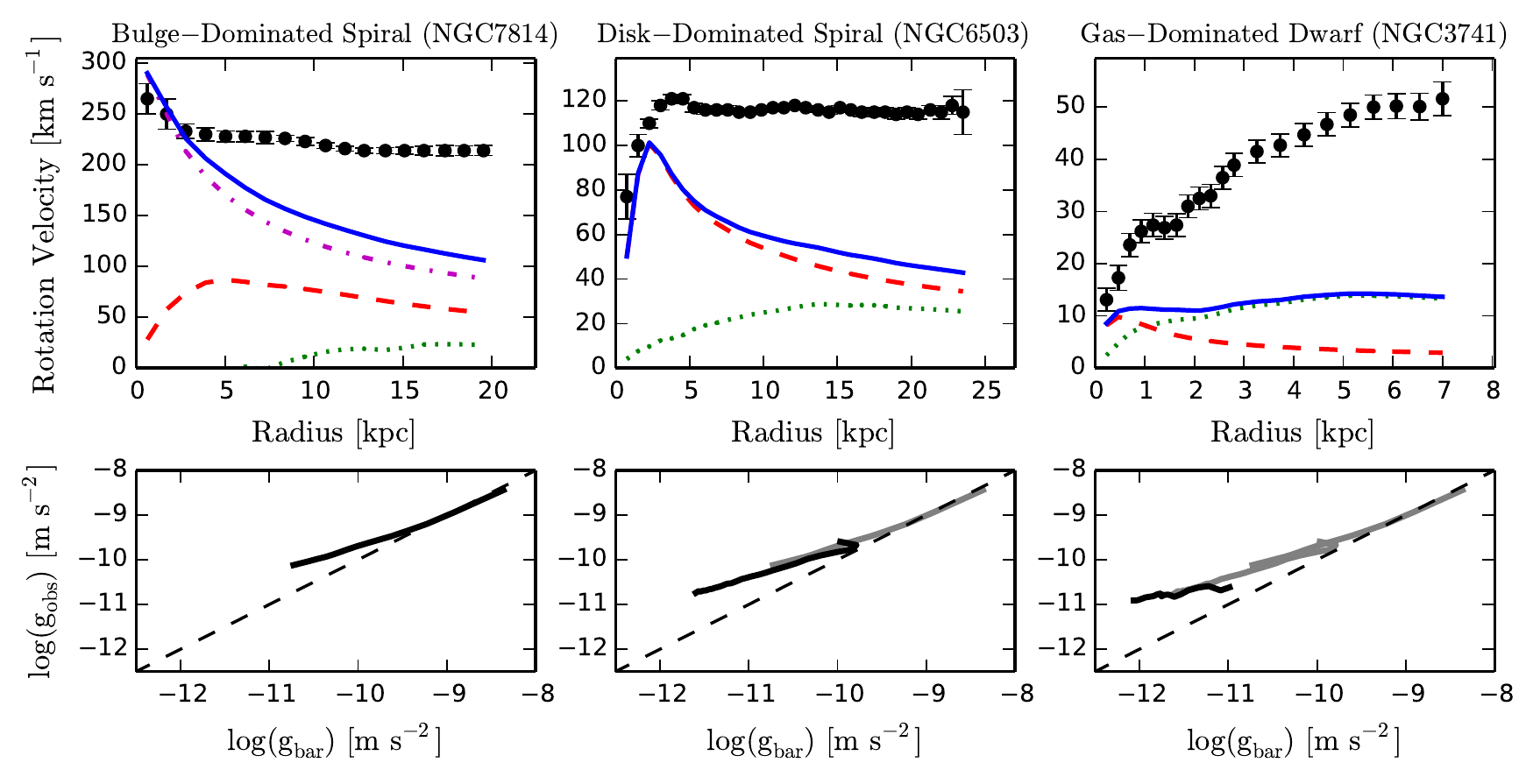}
\caption{Examples of mass models and rotation curves for individual galaxies.
The points with error bars in the upper panels are the observed rotation curves $V(R)$.
The errors represent both random errors and systematic uncertainty in the circular velocity 
due to asymmetry in the velocity field. 
In all galaxies, the data exceed the lines $v_{\mathrm{bar}} = \sqrt{R \gbar}$
representing the baryonic mass models (eq.~\ref{eq:gbar}), indicating the need for dark matter.
Each baryonic component is represented: 
dotted lines for the gas, dashed lines for the stellar disk, and dash-dotted lines for the bulge, when present.
The sum of these components is the baryonic mass model (solid line).
The lower panels illustrate the run of \gbar\ and \gobs\ for each galaxy, with the
dashed line being the line of unity.  {Note that higher accelerations occur at smaller radii.}
From left to right each line is replotted in gray to illustrate 
how {progressively fainter galaxies probe progressively lower} regimes of acceleration. 
\label{massmodels}}
\end{figure*}

We solve the Poisson equation 
\begin{equation}
\nabla^2 \Phi_{\mathrm{bar}} = 4 \pi G \rho_{\mathrm{bar}}
\label{eq:poisson}
\end{equation}
numerically \cite{Casertano,BT,GIPSY} to determine the gravitational potential $\Phi_{\mathrm{bar}}$ of 
each baryonic component (Fig.~\ref{massmodels}).
The acceleration due to the sum of baryonic components is 
\begin{equation}
\gbar =  \left| \frac{\partial \Phi_{\mathrm{bar}}}{\partial{R}} \right|.
\label{eq:gbar}
\end{equation}
Note that this refers only to the observed baryons. It is measured independently 
of the {actual acceleration} \gobs\ obtained from the rotation curve.

While the majority of stars and gas resides in thin disks, some galaxies have a central, quasi-spherical bulge component.
These bulges represent an important component of the stellar mass in only 31 of the 153 SPARC
galaxies. For these galaxies, we treat the bulges as spherical mass components distinct from the stellar 
disks. This detail only affects the the estimate of \gbar\ at the innermost points of a few 
galaxies with large bulges.

\subsection{Stellar Mass-to-Light Ratios}
\label{MLprescription}

We observe starlight while physics requires stellar mass. 
The mass-to-light ratio \ML\ is thus an unavoidable conversion factor.
The most robust indicator of stellar mass is the near-infrared luminosity \cite{BdJ}.

\begin{figure}
\includegraphics[width=3.5in]{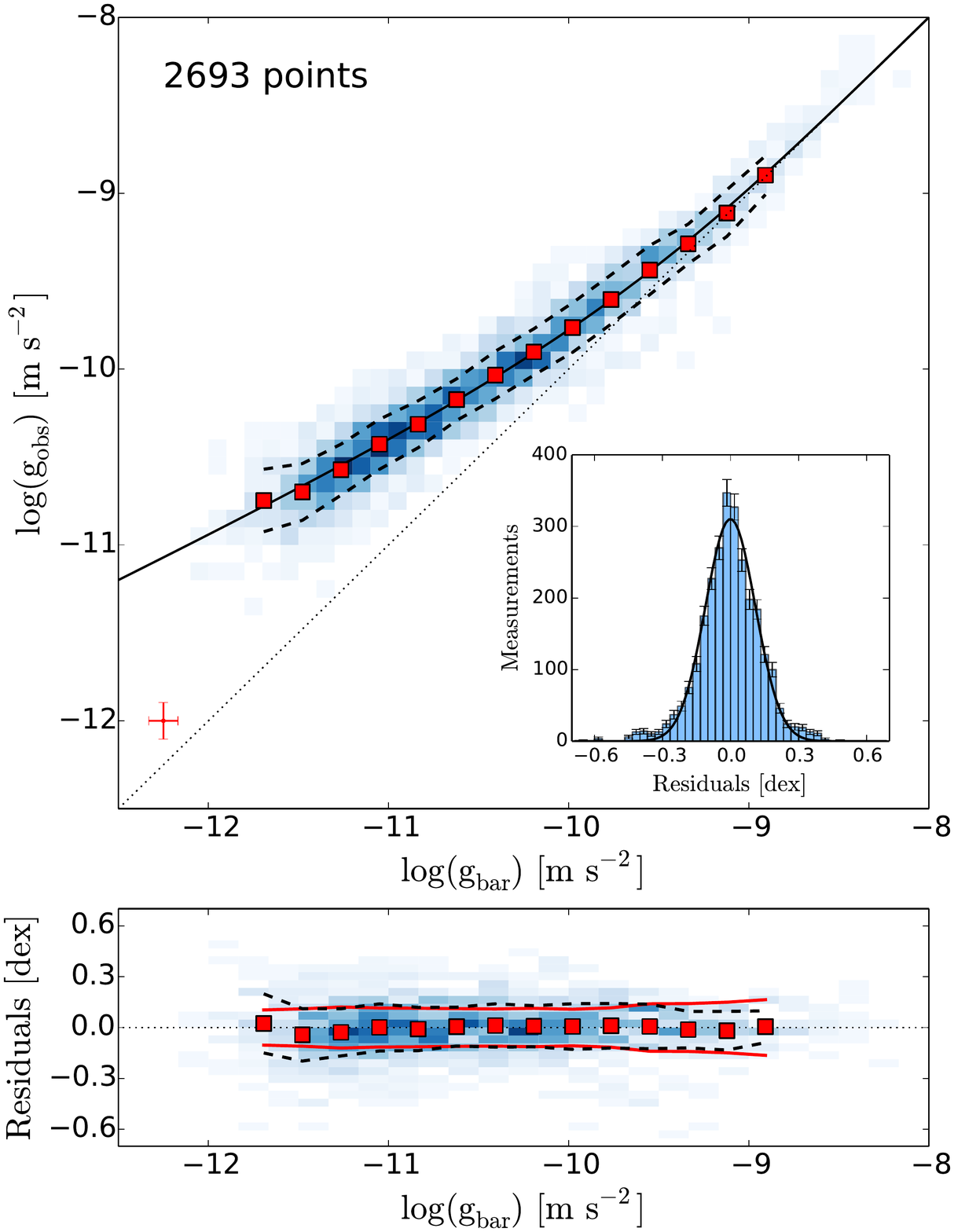}
\caption{The centripetal acceleration observed in rotation curves, $\gobs = V^2/R$, is plotted against that
predicted for the observed distribution of baryons, $\gbar = | \partial \Phi_{\mathrm{bar}} / \partial R |$ in the upper panel.
Nearly 2700 individual data points for 153 SPARC galaxies are shown in grayscale.
The mean uncertainty on individual points is illustrated in the lower left corner.  
Large squares show the mean of binned data. Dashed lines show the width of the ridge
as measured by the rms in each bin.  The dotted line is the line of unity. 
The solid line is the fit of eq.~\ref{eq:RFRfit} to the 
unbinned data using an orthogonal-distance-regression algorithm that considers errors on both variables.
The inset shows the histogram of all residuals and a Gaussian of width $\sigma = 0.11$ dex. 
The residuals are shown as a function of \gobs\ in the lower panel. 
The error bars on the binned data are smaller than the size of the points.
The solid lines show the scatter expected from observational uncertainties and
galaxy to galaxy variation in the stellar mass-to-light ratio.
This extrinsic scatter closely follows the observed rms scatter (dashed lines):
the data are consistent with negligible intrinsic scatter.
\label{RFR}}
\end{figure}

We have constructed stellar population synthesis models of star forming disk galaxies \cite{SMpaper3} to estimate 
the mass-to-light ratio in the 3.6$\mu$m band of \textit{Spitzer}.  
The numerical value of $\ML^{[3.6]}$ depends only weakly on age and metallicity for a broad range of
models with different star formation histories.
Here we adopt $\ML^{[3.6]} = 0.50\;\MLsun$ \cite{SMpaper3} as representative of all disks of all morphological types.
Independent estimates range from $0.42\;\MLsun$ \cite{MS14} to $0.60\;\MLsun$ \cite{meidt}. 
By astronomical standards, this is a small systematic uncertainty, which we explore in a companion paper \cite{SPARCII}.
Adopting different \ML\ only affects details, not the basic result.

The use of a single mass-to-light ratio is a great advance over previous work. 
Rather than treat \ML\ as an adjustable parameter for each
and every galaxy \cite{M04}, it is fixed to a single value for all disks.
While there is surely some scatter about the central value,
adopting a universal \ML\ provides a direct representation of the data with an absolute minimum of assumptions.
{It essentially just places the stars and gas on the same scale. The basic result follows simply from the luminosity
profiles of each component.}

We make one small concession to astronomical complexity.
While population synthesis models predict very similar \ML\ for all star forming disks,
they anticipate higher \ML\ for the old stars of central bulges. Hence we adopt  
$\ML^{[3.6]} = 0.7\;\MLsun$ for bulges \cite{SMpaper3}. 
This two-component population model only applies to the 31 of 153 galaxies with bulges,
and has only a small effect on
the estimate of \gbar\ in the innermost regions where the bulge dominates (Fig.~\ref{massmodels}).

\section{Results}

The mass models of individual galaxies are quite diverse (Fig.~\ref{massmodels}).
Bright, high surface brightness galaxies have stellar components that make a substantial contribution to the mass at small radii.
Indeed, it is common for these objects to approach the regime of ``maximum disk'' \cite{vAS1986}.  
Stars suffice to explain most of the observed rotation at small radii
(Fig.~\ref{massmodels}). 
At the opposite extreme, the mass discrepancy is large ($V \gg V_{\mathrm{bar}}$) in low surface brightness galaxies.
These require lots of dark matter, even at small radii \cite{MdB98a,dBM97}. Nevertheless,
the observed acceleration \gobs\ correlates strongly with that predicted by the baryons \gbar\ for all galaxies (Fig.~\ref{RFR}).

\begin{table}
\caption{\label{tab:table1} Scatter Budget for Acceleration Residuals}
\begin{ruledtabular}
\begin{tabular}{lr}
Source & Residual \\
\hline
Rotation velocity errors &Ê 0.03 dex  \\
Disk inclination errorsÊÊ &  0.05Ê dex  \\
Galaxy distanceÊ errors &Ê 0.08 dex  \\
Variation in mass-to-light ratios   & 0.06 dex \\
HI flux calibration errors  & 0.01 dex \\
\hline
TotalÊÊÊÊÊÊÊÊÊÊÊÊÊÊÊÊÊÊÊÊÊ & {0.12 dex} \\
\end{tabular}
\end{ruledtabular}
\end{table}

The correlation between \gobs\ and \gbar\ in Fig.~\ref{RFR} refers to the observed and expected centripetal acceleration.
Initially, this radial acceleration relation might seem trivial: acceleration correlates with acceleration.
However, the axes of Fig.~\ref{RFR} are completely independent. The ordinate, \gobs, is obtained from the rotation curves.
The abscissa, \gbar, is obtained from the observed distribution of baryons via the Poisson equation.
There is no guarantee that \gobs\ should correlate with \gbar\ when dark matter dominates.

Nevertheless, the radial acceleration relation persists for all galaxies of all types. Some galaxies only probe the high acceleration regime
while others only probe the low end (Fig.~\ref{massmodels}). The outer regions of high surface brightness
galaxies map smoothly to the inner regions of low surface brightness galaxies. These very different objects 
evince the same mass discrepancy at the same acceleration.
Individual galaxies are indistinguishable in Fig.~\ref{RFR}.

{Figure 3 combines and generalizes four well-established properties of rotating galaxies: flat rotation curves in the outer parts of 
spiral galaxies \cite{vera,bosma}; the ``conspiracy'' that spiral rotation curves show no indication of the transition from the baryon-dominated 
inner regions to the outer parts that are dark matter-dominated in the standard model \cite{vAS1986}; the Tully-Fisher \cite{TForig} relation 
between the outer velocity and the inner stellar mass, later generalized to the stellar plus 
atomic hydrogen mass \cite{btforig}; and the relation between the central surface brightness of galaxies and their inner rotation curve 
gradient \cite{dBM96,RCgradient,CentralDensRelation}.}

{It is convenient to fit a function that describes the data.}
The function \cite{M08}
\begin{equation}
\gobs = {\cal F}(\gbar) = \frac{\gbar}{1-e^{-\sqrt{\gbar/\azero}}}
\label{eq:RFRfit}
\end{equation}
provides a good fit.
The one fit parameter is the acceleration scale, \azero, where the mass discrepancy becomes pronounced.
For our adopted \ML, we find $\azero = 1.20 \pm 0.02$ (random) $\pm 0.24$ (systematic) $\times 10^{-10}\;\accunits$. 
The random error is a $1 \sigma$ value, while the systematic uncertainty represents the 20\% normalization 
uncertainty in \ML.

Equation \ref{eq:RFRfit} provides a good description of $\sim$2700 individual data points in 153 different galaxies.
This is a rather minimalistic parameterization. In addition to the scale \azero, eq.\ \ref{eq:RFRfit} implicitly contains a
linear slope at high accelerations and $\gobs \propto \sqrt{\gbar}$ at low accelerations. The high end slope is sensible:
dark matter becomes negligible at some point. The low end slope of the data could in principle differ from that implicitly
assumed by eq.\ \ref{eq:RFRfit}, but if so there is no indication in these data.

{Residuals from the fit are well described by a Gaussian of width 0.11 dex (Fig.\ \ref{RFR}). 
The rms scatter is 0.13 dex owing to the inevitable outliers.  These are tiny numbers by the standards of extragalactic astronomy.
The intrinsic scatter in the relation must be smaller still once scatter due to errors are accounted for.}

There are two {types of extrinsic} scatter in the radial acceleration relation: measurement uncertainties and galaxy to galaxy variation in \ML. 
Measurement uncertainties in \gobs\ follow from the error in the rotation velocities, disk inclinations, and galaxy distances. 
{The mean contribution of each is given in Table \ref{tab:table1}}.
Intrinsic scatter about the mean mass-to-light ratio is anticipated to be 0.11 dex at 3.6$\mu$m \cite{meidt}.
{This propagates to a net residual of 0.06 dex in \gbar\ after accounting for the variable slope of the relation.
The total expected scatter is 0.12 dex (Table \ref{tab:table1}), leaving little room for intrinsic scatter.}

{Astronomical data often suffer from unrecognized systematics. In the case of rotation curves, this is frequently
argued \cite{vdBSwat,OmanApostle,Pinhead} to be the cause of the apparent discrepancy \cite{dBcorecusp} with the predictions of 
numerical simulations \cite{NFW}.  This cannot be the case here.
If we had neglected some important source of uncertainty, we would erroneously infer a large intrinsic scatter, not a small one. 
For the intrinsic scatter to be non-negligible, the errors must be overestimated rather than underestimated.
If there were no observational uncertainty at all, the intrinsic scatter would still be limited by the small observed rms of 0.13 dex.}

{Regardless of whether the intrinsic scatter is zero or merely very small, the radial acceleration relation is an important empirical
facet of the mass discrepancy problem.} 
When \gbar\ is observed, \gobs\ follows, and vice-versa.
{This must be explained by any successful theory.}

\section{Discussion}

We find a strong relation between the observed radial acceleration \gobs\ and that due to the baryons, \gbar.
This radial acceleration relation is completely empirical.  It follows from a minimum of assumptions.
The only inputs are the data, the Poisson equation, and the simplest possible conversion of 
starlight to stellar mass.

We have not considered any particular halo model for the dark matter.  Indeed, such models are unnecessary.
The distribution of dark matter follows directly from the relation, and can be written entirely in terms of the baryons:
\begin{equation}
\mathrm{g}_{\mathrm{DM}} = \gobs - \gbar = \frac{\gbar}{e^{\sqrt{\gbar/\azero}} -1}.
\label{eq:DMdist}
\end{equation}
The dark and baryonic mass are strongly coupled \cite{M04,renzorule}.

Possible interpretations for the radial acceleration relation fall into three broad categories.
\begin{enumerate}
\item It represents the end product of galaxy formation.
\item It represents new dark sector physics that leads to the observed coupling.
\item It is the result of new dynamical laws rather than dark matter.
\end{enumerate}
None of these options are entirely satisfactory.

In the standard cosmological paradigm, galaxies form within dark matter halos.
Simulations of this process do not naturally lead to realistic galaxies \cite{dBcorecusp,McG2014}.
Complicated accessory effects (``feedback'') must be invoked to remodel simulated
galaxies into something more akin to observations. {Whether such processes can satisfactorily explain
the radial acceleration relation and its small scatter remains to be demonstrated \cite{DCL,Desmond}.}

Another possibility is new ``dark sector'' physics.
The dark matter needs to respond to the distribution of baryons
(or vice-versa) in order to give the observed relation. This is not trivial to achieve, but 
the observed phenomenology might emerge if dark matter behaves as a fluid \cite{darkfluid,Khoury2015} or 
is subject to gravitational polarization \cite{blanchet}. 

Thirdly, the one-to-one correspondence between \gbar\ and \gobs\ suggests that the baryons are the source of the gravitational potential.
In this case, one might alter the laws of dynamics rather than invoke dark matter. 
Indeed, our results were anticipated over three decades ago by MOND \cite{milgrom83}. 
Whether this is a situation in which it would be necessary to invent MOND if it did not already exist is worthy of contemplation.

In MOND, eq.\ \ref{eq:RFRfit} is related to the MOND interpolation function.
However, we should be careful not to confuse data with theory. 
Equation \ref{eq:RFRfit} provides a convenient description of the data irrespective of MOND.


Regardless of its theoretical basis, the radial acceleration relation exists as an empirical relation.
The acceleration scale \azero\ is in the data. 
The observed coupling between \gobs\ and \gbar\ demands a satisfactory explanation.
The radial acceleration relation appears to be a law of nature, a sort of Kepler's law for rotating galaxies. \\

\begin{acknowledgements}
We thank the referees and editorial staff for their thorough and diligent attention. 
This work would not be possible without the efforts of many dozens of observers working at both radio and
optical wavelengths over the past several decades; in particular the many Ph.D.\ students at the University
of Groningen trained by Profs.\ van Albada and Sancisi.
We are also grateful to Jim Peebles and David Merritt for perspective and encouragement.
This work is based in part on observations made with the Spitzer Space Telescope, which is operated by the 
Jet Propulsion Laboratory, California Institute of Technology under a contract with NASA.
This publication was made possible through the support of the John Templeton Foundation. 
The opinions expressed here are those of the authors and do not necessary reflect the views of the John Templeton Foundation.
\end{acknowledgements}


\begin{thebibliography}{52}%
\makeatletter
\providecommand \@ifxundefined [1]{%
 \@ifx{#1\undefined}
}%
\providecommand \@ifnum [1]{%
 \ifnum #1\expandafter \@firstoftwo
 \else \expandafter \@secondoftwo
 \fi
}%
\providecommand \@ifx [1]{%
 \ifx #1\expandafter \@firstoftwo
 \else \expandafter \@secondoftwo
 \fi
}%
\providecommand \natexlab [1]{#1}%
\providecommand \enquote  [1]{``#1''}%
\providecommand \bibnamefont  [1]{#1}%
\providecommand \bibfnamefont [1]{#1}%
\providecommand \citenamefont [1]{#1}%
\providecommand \href@noop [0]{\@secondoftwo}%
\providecommand \href [0]{\begingroup \@sanitize@url \@href}%
\providecommand \@href[1]{\@@startlink{#1}\@@href}%
\providecommand \@@href[1]{\endgroup#1\@@endlink}%
\providecommand \@sanitize@url [0]{\catcode `\\12\catcode `\$12\catcode
  `\&12\catcode `\#12\catcode `\^12\catcode `\_12\catcode `\%12\relax}%
\providecommand \@@startlink[1]{}%
\providecommand \@@endlink[0]{}%
\providecommand \url  [0]{\begingroup\@sanitize@url \@url }%
\providecommand \@url [1]{\endgroup\@href {#1}{\urlprefix }}%
\providecommand \urlprefix  [0]{URL }%
\providecommand \Eprint [0]{\href }%
\providecommand \doibase [0]{http://dx.doi.org/}%
\providecommand \selectlanguage [0]{\@gobble}%
\providecommand \bibinfo  [0]{\@secondoftwo}%
\providecommand \bibfield  [0]{\@secondoftwo}%
\providecommand \translation [1]{[#1]}%
\providecommand \BibitemOpen [0]{}%
\providecommand \bibitemStop [0]{}%
\providecommand \bibitemNoStop [0]{.\EOS\space}%
\providecommand \EOS [0]{\spacefactor3000\relax}%
\providecommand \BibitemShut  [1]{\csname bibitem#1\endcsname}%
\let\auto@bib@innerbib\@empty
\bibitem [{\citenamefont {{Rubin}}\ \emph {et~al.}(1978)\citenamefont
  {{Rubin}}, \citenamefont {{Thonnard}},\ and\ \citenamefont {{Ford}}}]{vera}%
  \BibitemOpen
  \bibfield  {author} {\bibinfo {author} {\bibfnamefont {V.~C.}\ \bibnamefont
  {{Rubin}}}, \bibinfo {author} {\bibfnamefont {N.}~\bibnamefont {{Thonnard}}},
  \ and\ \bibinfo {author} {\bibfnamefont {W.~K.}\ \bibnamefont {{Ford}},
  \bibfnamefont {Jr.}},\ }\href {\doibase 10.1086/182804} {\bibfield  {journal}
  {\bibinfo  {journal} {Astrophys. J.}\ }\textbf {\bibinfo {volume} {225}},\
  \bibinfo {pages} {L107} (\bibinfo {year} {1978})}\BibitemShut {NoStop}%
\bibitem [{\citenamefont {{Bosma}}(1981)}]{bosma}%
  \BibitemOpen
  \bibfield  {author} {\bibinfo {author} {\bibfnamefont {A.}~\bibnamefont
  {{Bosma}}},\ }\href {\doibase 10.1086/113062} {\bibfield  {journal} {\bibinfo
   {journal} {Astron. J.}\ }\textbf {\bibinfo {volume} {86}},\ \bibinfo {pages}
  {1791} (\bibinfo {year} {1981})}\BibitemShut {NoStop}%
\bibitem [{\citenamefont {{Tully}}\ and\ \citenamefont
  {{Fisher}}(1977)}]{TForig}%
  \BibitemOpen
  \bibfield  {author} {\bibinfo {author} {\bibfnamefont {R.~B.}\ \bibnamefont
  {{Tully}}}\ and\ \bibinfo {author} {\bibfnamefont {J.~R.}\ \bibnamefont
  {{Fisher}}},\ }\href@noop {} {\bibfield  {journal} {\bibinfo  {journal}
  {Astron. Astrophys.}\ }\textbf {\bibinfo {volume} {54}},\ \bibinfo {pages}
  {661} (\bibinfo {year} {1977})}\BibitemShut {NoStop}%
\bibitem [{\citenamefont {{McGaugh}}\ \emph {et~al.}(2000)\citenamefont
  {{McGaugh}}, \citenamefont {{Schombert}}, \citenamefont {{Bothun}},\ and\
  \citenamefont {{de Blok}}}]{btforig}%
  \BibitemOpen
  \bibfield  {author} {\bibinfo {author} {\bibfnamefont {S.~S.}\ \bibnamefont
  {{McGaugh}}}, \bibinfo {author} {\bibfnamefont {J.~M.}\ \bibnamefont
  {{Schombert}}}, \bibinfo {author} {\bibfnamefont {G.~D.}\ \bibnamefont
  {{Bothun}}}, \ and\ \bibinfo {author} {\bibfnamefont {W.~J.~G.}\ \bibnamefont
  {{de Blok}}},\ }\href {\doibase 10.1086/312628} {\bibfield  {journal}
  {\bibinfo  {journal} {Astrophys. J.}\ }\textbf {\bibinfo {volume} {533}},\
  \bibinfo {pages} {L99} (\bibinfo {year} {2000})}\BibitemShut {NoStop}%
\bibitem [{\citenamefont {{Verheijen}}(2001)}]{verhTF}%
  \BibitemOpen
  \bibfield  {author} {\bibinfo {author} {\bibfnamefont {M.~A.~W.}\
  \bibnamefont {{Verheijen}}},\ }\href@noop {} {\bibfield  {journal} {\bibinfo
  {journal} {Astrophys. J.}\ }\textbf {\bibinfo {volume} {563}},\ \bibinfo
  {pages} {694} (\bibinfo {year} {2001})}\BibitemShut {NoStop}%
\bibitem [{\citenamefont {{Courteau}}\ and\ \citenamefont {{Rix}}(1999)}]{CR}%
  \BibitemOpen
  \bibfield  {author} {\bibinfo {author} {\bibfnamefont {S.}~\bibnamefont
  {{Courteau}}}\ and\ \bibinfo {author} {\bibfnamefont {H.}~\bibnamefont
  {{Rix}}},\ }\href {\doibase 10.1086/306872} {\bibfield  {journal} {\bibinfo
  {journal} {Astrophys. J.}\ }\textbf {\bibinfo {volume} {513}},\ \bibinfo
  {pages} {561} (\bibinfo {year} {1999})}\BibitemShut {NoStop}%
\bibitem [{\citenamefont {{McGaugh}}(2005)}]{myPRL}%
  \BibitemOpen
  \bibfield  {author} {\bibinfo {author} {\bibfnamefont {S.~S.}\ \bibnamefont
  {{McGaugh}}},\ }\href {\doibase 10.1103/PhysRevLett.95.171302} {\bibfield
  {journal} {\bibinfo  {journal} {Phys. Rev. Lett.}\ }\textbf {\bibinfo
  {volume} {95}},\ \bibinfo {pages} {171302} (\bibinfo {year}
  {2005})}\BibitemShut {NoStop}%
\bibitem [{\citenamefont {{Zwaan}}\ \emph {et~al.}(1995)\citenamefont
  {{Zwaan}}, \citenamefont {{van der Hulst}}, \citenamefont {{de Blok}},\ and\
  \citenamefont {{McGaugh}}}]{zwaanTF}%
  \BibitemOpen
  \bibfield  {author} {\bibinfo {author} {\bibfnamefont {M.~A.}\ \bibnamefont
  {{Zwaan}}}, \bibinfo {author} {\bibfnamefont {J.~M.}\ \bibnamefont {{van der
  Hulst}}}, \bibinfo {author} {\bibfnamefont {W.~J.~G.}\ \bibnamefont {{de
  Blok}}}, \ and\ \bibinfo {author} {\bibfnamefont {S.~S.}\ \bibnamefont
  {{McGaugh}}},\ }\href@noop {} {\bibfield  {journal} {\bibinfo  {journal}
  {Mon. Not. R. Astron. Soc.}\ }\textbf {\bibinfo {volume} {273}},\ \bibinfo
  {pages} {L35} (\bibinfo {year} {1995})}\BibitemShut {NoStop}%
\bibitem [{\citenamefont {{McGaugh}}\ and\ \citenamefont {{de
  Blok}}(1998)}]{MdB98a}%
  \BibitemOpen
  \bibfield  {author} {\bibinfo {author} {\bibfnamefont {S.~S.}\ \bibnamefont
  {{McGaugh}}}\ and\ \bibinfo {author} {\bibfnamefont {W.~J.~G.}\ \bibnamefont
  {{de Blok}}},\ }\href@noop {} {\bibfield  {journal} {\bibinfo  {journal}
  {Astrophys. J.}\ }\textbf {\bibinfo {volume} {499}},\ \bibinfo {pages} {41}
  (\bibinfo {year} {1998})}\BibitemShut {NoStop}%
\bibitem [{\citenamefont {{McGaugh}}(2011)}]{M11}%
  \BibitemOpen
  \bibfield  {author} {\bibinfo {author} {\bibfnamefont {S.~S.}\ \bibnamefont
  {{McGaugh}}},\ }\href {\doibase 10.1103/PhysRevLett.106.121303} {\bibfield
  {journal} {\bibinfo  {journal} {Physical Review Letters}\ }\textbf {\bibinfo
  {volume} {106}},\ \bibinfo {eid} {121303} (\bibinfo {year} {2011})},\ \Eprint
  {http://arxiv.org/abs/1102.3913} {arXiv:1102.3913 [astro-ph.CO]} \BibitemShut
  {NoStop}%
\bibitem [{\citenamefont {{McGaugh}}\ and\ \citenamefont
  {{Schombert}}(2015)}]{MS15}%
  \BibitemOpen
  \bibfield  {author} {\bibinfo {author} {\bibfnamefont {S.~S.}\ \bibnamefont
  {{McGaugh}}}\ and\ \bibinfo {author} {\bibfnamefont {J.~M.}\ \bibnamefont
  {{Schombert}}},\ }\href {\doibase 10.1088/0004-637X/802/1/18} {\bibfield
  {journal} {\bibinfo  {journal} {Astrophys. J.}\ }\textbf {\bibinfo {volume}
  {802}},\ \bibinfo {eid} {18} (\bibinfo {year} {2015})},\ \Eprint
  {http://arxiv.org/abs/1501.06826} {arXiv:1501.06826} \BibitemShut {NoStop}%
\bibitem [{\citenamefont {{Lelli}}\ \emph
  {et~al.}(2016{\natexlab{a}})\citenamefont {{Lelli}}, \citenamefont
  {{McGaugh}},\ and\ \citenamefont {{Schombert}}}]{LMS2016}%
  \BibitemOpen
  \bibfield  {author} {\bibinfo {author} {\bibfnamefont {F.}~\bibnamefont
  {{Lelli}}}, \bibinfo {author} {\bibfnamefont {S.~S.}\ \bibnamefont
  {{McGaugh}}}, \ and\ \bibinfo {author} {\bibfnamefont {J.~M.}\ \bibnamefont
  {{Schombert}}},\ }\href {\doibase 10.3847/2041-8205/816/1/L14} {\bibfield
  {journal} {\bibinfo  {journal} {Astrophys. J.}\ }\textbf {\bibinfo {volume}
  {816}},\ \bibinfo {eid} {L14} (\bibinfo {year} {2016}{\natexlab{a}})},\
  \Eprint {http://arxiv.org/abs/1512.04543} {arXiv:1512.04543} \BibitemShut
  {NoStop}%
\bibitem [{\citenamefont {{Sancisi}}(2004)}]{renzorule}%
  \BibitemOpen
  \bibfield  {author} {\bibinfo {author} {\bibfnamefont {R.}~\bibnamefont
  {{Sancisi}}},\ }in\ \href@noop {} {\emph {\bibinfo {booktitle} {Dark Matter
  in Galaxies}}},\ \bibinfo {series} {IAU Symposium}, Vol.\ \bibinfo {volume}
  {220},\ \bibinfo {editor} {edited by\ \bibinfo {editor} {\bibfnamefont
  {S.}~\bibnamefont {{Ryder}}}, \bibinfo {editor} {\bibfnamefont
  {D.}~\bibnamefont {{Pisano}}}, \bibinfo {editor} {\bibfnamefont
  {M.}~\bibnamefont {{Walker}}}, \ and\ \bibinfo {editor} {\bibfnamefont
  {K.}~\bibnamefont {{Freeman}}}}\ (\bibinfo {year} {2004})\ p.\ \bibinfo
  {pages} {233},\ \Eprint {http://arxiv.org/abs/arXiv:astro-ph/0311348}
  {arXiv:astro-ph/0311348} \BibitemShut {NoStop}%
\bibitem [{\citenamefont {{McGaugh}}(2004)}]{M04}%
  \BibitemOpen
  \bibfield  {author} {\bibinfo {author} {\bibfnamefont {S.~S.}\ \bibnamefont
  {{McGaugh}}},\ }\href {\doibase 10.1086/421338} {\bibfield  {journal}
  {\bibinfo  {journal} {Astrophys. J.}\ }\textbf {\bibinfo {volume} {609}},\
  \bibinfo {pages} {652} (\bibinfo {year} {2004})}\BibitemShut {NoStop}%
\bibitem [{\citenamefont {{McGaugh}}(2014)}]{GalRev}%
  \BibitemOpen
  \bibfield  {author} {\bibinfo {author} {\bibfnamefont {S.~S.}\ \bibnamefont
  {{McGaugh}}},\ }\href {\doibase 10.3390/galaxies2040601} {\bibfield
  {journal} {\bibinfo  {journal} {Galaxies}\ }\textbf {\bibinfo {volume} {2}},\
  \bibinfo {pages} {601} (\bibinfo {year} {2014})},\ \Eprint
  {http://arxiv.org/abs/1412.3767} {arXiv:1412.3767} \BibitemShut {NoStop}%
\bibitem [{\citenamefont {{Scarpa}}(2006)}]{Scarpa2006}%
  \BibitemOpen
  \bibfield  {author} {\bibinfo {author} {\bibfnamefont {R.}~\bibnamefont
  {{Scarpa}}},\ }in\ \href {\doibase 10.1063/1.2189141} {\emph {\bibinfo
  {booktitle} {First Crisis in Cosmology Conference}}},\ \bibinfo {series}
  {American Institute of Physics Conference Series}, Vol.\ \bibinfo {volume}
  {822},\ \bibinfo {editor} {edited by\ \bibinfo {editor} {\bibfnamefont
  {E.~J.}\ \bibnamefont {{Lerner}}}\ and\ \bibinfo {editor} {\bibfnamefont
  {J.~B.}\ \bibnamefont {{Almeida}}}}\ (\bibinfo {year} {2006})\ pp.\ \bibinfo
  {pages} {253--265},\ \Eprint {http://arxiv.org/abs/astro-ph/0601478}
  {astro-ph/0601478} \BibitemShut {NoStop}%
\bibitem [{\citenamefont {{den Heijer}}\ \emph {et~al.}(2015)\citenamefont
  {{den Heijer}}, \citenamefont {{Oosterloo}}, \citenamefont {{Serra}},
  \citenamefont {{J{\'o}zsa}}, \citenamefont {{Kerp}}, \citenamefont
  {{Morganti}}, \citenamefont {{Cappellari}}, \citenamefont {{Davis}},
  \citenamefont {{Duc}}, \citenamefont {{Emsellem}}, \citenamefont
  {{Krajnovi{\'c}}}, \citenamefont {{McDermid}}, \citenamefont {{Naab}},
  \citenamefont {{Weijmans}},\ and\ \citenamefont {{de
  Zeeuw}}}]{ellipticalBTF}%
  \BibitemOpen
  \bibfield  {author} {\bibinfo {author} {\bibfnamefont {M.}~\bibnamefont {{den
  Heijer}}}, \bibinfo {author} {\bibfnamefont {T.~A.}\ \bibnamefont
  {{Oosterloo}}}, \bibinfo {author} {\bibfnamefont {P.}~\bibnamefont
  {{Serra}}}, \bibinfo {author} {\bibfnamefont {G.~I.~G.}\ \bibnamefont
  {{J{\'o}zsa}}}, \bibinfo {author} {\bibfnamefont {J.}~\bibnamefont {{Kerp}}},
  \bibinfo {author} {\bibfnamefont {R.}~\bibnamefont {{Morganti}}}, \bibinfo
  {author} {\bibfnamefont {M.}~\bibnamefont {{Cappellari}}}, \bibinfo {author}
  {\bibfnamefont {T.~A.}\ \bibnamefont {{Davis}}}, \bibinfo {author}
  {\bibfnamefont {P.-A.}\ \bibnamefont {{Duc}}}, \bibinfo {author}
  {\bibfnamefont {E.}~\bibnamefont {{Emsellem}}}, \bibinfo {author}
  {\bibfnamefont {D.}~\bibnamefont {{Krajnovi{\'c}}}}, \bibinfo {author}
  {\bibfnamefont {R.~M.}\ \bibnamefont {{McDermid}}}, \bibinfo {author}
  {\bibfnamefont {T.}~\bibnamefont {{Naab}}}, \bibinfo {author} {\bibfnamefont
  {A.-M.}\ \bibnamefont {{Weijmans}}}, \ and\ \bibinfo {author} {\bibfnamefont
  {P.~T.}\ \bibnamefont {{de Zeeuw}}},\ }\href {\doibase
  10.1051/0004-6361/201526879} {\bibfield  {journal} {\bibinfo  {journal}
  {Astron. Astrophys.}\ }\textbf {\bibinfo {volume} {581}},\ \bibinfo {eid}
  {A98} (\bibinfo {year} {2015})},\ \Eprint {http://arxiv.org/abs/1509.05236}
  {arXiv:1509.05236} \BibitemShut {NoStop}%
\bibitem [{\citenamefont {{Serra}}\ \emph {et~al.}(2016)\citenamefont
  {{Serra}}, \citenamefont {{Oosterloo}}, \citenamefont {{Cappellari}},
  \citenamefont {{den Heijer}},\ and\ \citenamefont {{J{\'o}zsa}}}]{Serra2016}%
  \BibitemOpen
  \bibfield  {author} {\bibinfo {author} {\bibfnamefont {P.}~\bibnamefont
  {{Serra}}}, \bibinfo {author} {\bibfnamefont {T.}~\bibnamefont
  {{Oosterloo}}}, \bibinfo {author} {\bibfnamefont {M.}~\bibnamefont
  {{Cappellari}}}, \bibinfo {author} {\bibfnamefont {M.}~\bibnamefont {{den
  Heijer}}}, \ and\ \bibinfo {author} {\bibfnamefont {G.~I.~G.}\ \bibnamefont
  {{J{\'o}zsa}}},\ }\href {\doibase 10.1093/mnras/stw1010} {\bibfield
  {journal} {\bibinfo  {journal} {Mon. Not. Royal Astron. Soc.}\ } (\bibinfo
  {year} {2016}),\ 10.1093/mnras/stw1010}\BibitemShut {NoStop}%
\bibitem [{\citenamefont {{Lelli}}\ \emph
  {et~al.}(2016{\natexlab{b}})\citenamefont {{Lelli}}, \citenamefont
  {{McGaugh}},\ and\ \citenamefont {{Schombert}}}]{SPARCI}%
  \BibitemOpen
  \bibfield  {author} {\bibinfo {author} {\bibfnamefont {F.}~\bibnamefont
  {{Lelli}}}, \bibinfo {author} {\bibfnamefont {S.~S.}\ \bibnamefont
  {{McGaugh}}}, \ and\ \bibinfo {author} {\bibfnamefont {J.~M.}\ \bibnamefont
  {{Schombert}}},\ }\href@noop {} {\bibfield  {journal} {\bibinfo  {journal}
  {Astron. J.}\ } (\bibinfo {year} {2016}{\natexlab{b}})},\ \Eprint
  {http://arxiv.org/abs/1606.09251} {arXiv:1606.09251} \BibitemShut {NoStop}%
\bibitem [{\citenamefont {{Lange}}\ \emph {et~al.}(2015)\citenamefont
  {{Lange}}, \citenamefont {{Driver}}, \citenamefont {{Robotham}},
  \citenamefont {{Kelvin}}, \citenamefont {{Graham}}, \citenamefont
  {{Alpaslan}}, \citenamefont {{Andrews}}, \citenamefont {{Baldry}},
  \citenamefont {{Bamford}}, \citenamefont {{Bland-Hawthorn}}, \citenamefont
  {{Brough}}, \citenamefont {{Cluver}}, \citenamefont {{Conselice}},
  \citenamefont {{Davies}}, \citenamefont {{Haeussler}}, \citenamefont
  {{Konstantopoulos}}, \citenamefont {{Loveday}}, \citenamefont {{Moffett}},
  \citenamefont {{Norberg}}, \citenamefont {{Phillipps}}, \citenamefont
  {{Taylor}}, \citenamefont {{L{\'o}pez-S{\'a}nchez}},\ and\ \citenamefont
  {{Wilkins}}}]{GAMA}%
  \BibitemOpen
  \bibfield  {author} {\bibinfo {author} {\bibfnamefont {R.}~\bibnamefont
  {{Lange}}}, \bibinfo {author} {\bibfnamefont {S.~P.}\ \bibnamefont
  {{Driver}}}, \bibinfo {author} {\bibfnamefont {A.~S.~G.}\ \bibnamefont
  {{Robotham}}}, \bibinfo {author} {\bibfnamefont {L.~S.}\ \bibnamefont
  {{Kelvin}}}, \bibinfo {author} {\bibfnamefont {A.~W.}\ \bibnamefont
  {{Graham}}}, \bibinfo {author} {\bibfnamefont {M.}~\bibnamefont
  {{Alpaslan}}}, \bibinfo {author} {\bibfnamefont {S.~K.}\ \bibnamefont
  {{Andrews}}}, \bibinfo {author} {\bibfnamefont {I.~K.}\ \bibnamefont
  {{Baldry}}}, \bibinfo {author} {\bibfnamefont {S.}~\bibnamefont {{Bamford}}},
  \bibinfo {author} {\bibfnamefont {J.}~\bibnamefont {{Bland-Hawthorn}}},
  \bibinfo {author} {\bibfnamefont {S.}~\bibnamefont {{Brough}}}, \bibinfo
  {author} {\bibfnamefont {M.~E.}\ \bibnamefont {{Cluver}}}, \bibinfo {author}
  {\bibfnamefont {C.~J.}\ \bibnamefont {{Conselice}}}, \bibinfo {author}
  {\bibfnamefont {L.~J.~M.}\ \bibnamefont {{Davies}}}, \bibinfo {author}
  {\bibfnamefont {B.}~\bibnamefont {{Haeussler}}}, \bibinfo {author}
  {\bibfnamefont {I.~S.}\ \bibnamefont {{Konstantopoulos}}}, \bibinfo {author}
  {\bibfnamefont {J.}~\bibnamefont {{Loveday}}}, \bibinfo {author}
  {\bibfnamefont {A.~J.}\ \bibnamefont {{Moffett}}}, \bibinfo {author}
  {\bibfnamefont {P.}~\bibnamefont {{Norberg}}}, \bibinfo {author}
  {\bibfnamefont {S.}~\bibnamefont {{Phillipps}}}, \bibinfo {author}
  {\bibfnamefont {E.~N.}\ \bibnamefont {{Taylor}}}, \bibinfo {author}
  {\bibfnamefont {{\'A}.~R.}\ \bibnamefont {{L{\'o}pez-S{\'a}nchez}}}, \ and\
  \bibinfo {author} {\bibfnamefont {S.~M.}\ \bibnamefont {{Wilkins}}},\ }\href
  {\doibase 10.1093/mnras/stu2467} {\bibfield  {journal} {\bibinfo  {journal}
  {Mon. Not. Royal Astron. Soc.}\ }\textbf {\bibinfo {volume} {447}},\ \bibinfo
  {pages} {2603} (\bibinfo {year} {2015})},\ \Eprint
  {http://arxiv.org/abs/1411.6355} {arXiv:1411.6355} \BibitemShut {NoStop}%
\bibitem [{\citenamefont {{Bradford}}\ \emph {et~al.}(2015)\citenamefont
  {{Bradford}}, \citenamefont {{Geha}},\ and\ \citenamefont
  {{Blanton}}}]{Bradford2015}%
  \BibitemOpen
  \bibfield  {author} {\bibinfo {author} {\bibfnamefont {J.~D.}\ \bibnamefont
  {{Bradford}}}, \bibinfo {author} {\bibfnamefont {M.~C.}\ \bibnamefont
  {{Geha}}}, \ and\ \bibinfo {author} {\bibfnamefont {M.~R.}\ \bibnamefont
  {{Blanton}}},\ }\href {\doibase 10.1088/0004-637X/809/2/146} {\bibfield
  {journal} {\bibinfo  {journal} {Astrophys. J.}\ }\textbf {\bibinfo {volume}
  {809}},\ \bibinfo {eid} {146} (\bibinfo {year} {2015})},\ \Eprint
  {http://arxiv.org/abs/1505.04819} {arXiv:1505.04819} \BibitemShut {NoStop}%
\bibitem [{\citenamefont {{Eckert}}\ \emph {et~al.}(2016)\citenamefont
  {{Eckert}}, \citenamefont {{Kannappan}}, \citenamefont {{Stark}},
  \citenamefont {{Moffett}}, \citenamefont {{Berlind}},\ and\ \citenamefont
  {{Norris}}}]{RESOLVE}%
  \BibitemOpen
  \bibfield  {author} {\bibinfo {author} {\bibfnamefont {K.~D.}\ \bibnamefont
  {{Eckert}}}, \bibinfo {author} {\bibfnamefont {S.~J.}\ \bibnamefont
  {{Kannappan}}}, \bibinfo {author} {\bibfnamefont {D.~V.}\ \bibnamefont
  {{Stark}}}, \bibinfo {author} {\bibfnamefont {A.~J.}\ \bibnamefont
  {{Moffett}}}, \bibinfo {author} {\bibfnamefont {A.~A.}\ \bibnamefont
  {{Berlind}}}, \ and\ \bibinfo {author} {\bibfnamefont {M.~A.}\ \bibnamefont
  {{Norris}}},\ }\href {\doibase 10.3847/0004-637X/824/2/124} {\bibfield
  {journal} {\bibinfo  {journal} {Astrophys. J.}\ }\textbf {\bibinfo {volume}
  {824}},\ \bibinfo {eid} {124} (\bibinfo {year} {2016})},\ \Eprint
  {http://arxiv.org/abs/1604.03957} {arXiv:1604.03957} \BibitemShut {NoStop}%
\bibitem [{\citenamefont {{Schombert}}\ and\ \citenamefont
  {{McGaugh}}(2014{\natexlab{a}})}]{SMpaper4}%
  \BibitemOpen
  \bibfield  {author} {\bibinfo {author} {\bibfnamefont {J.~M.}\ \bibnamefont
  {{Schombert}}}\ and\ \bibinfo {author} {\bibfnamefont {S.}~\bibnamefont
  {{McGaugh}}},\ }\href {\doibase 10.1017/pasa.2014.2} {\bibfield  {journal}
  {\bibinfo  {journal} {Pub. Astron. Soc. Australia}\ }\textbf {\bibinfo
  {volume} {31}},\ \bibinfo {eid} {e011} (\bibinfo {year}
  {2014}{\natexlab{a}})},\ \Eprint {http://arxiv.org/abs/1401.0238}
  {arXiv:1401.0238} \BibitemShut {NoStop}%
\bibitem [{\citenamefont {{Meidt}}\ \emph {et~al.}(2014)\citenamefont
  {{Meidt}}, \citenamefont {{Schinnerer}}, \citenamefont {{van de Ven}},
  \citenamefont {{Zaritsky}}, \citenamefont {{Peletier}}, \citenamefont
  {{Knapen}}, \citenamefont {{Sheth}}, \citenamefont {{Regan}}, \citenamefont
  {{Querejeta}}, \citenamefont {{Mu{\~n}oz-Mateos}}, \citenamefont {{Kim}},
  \citenamefont {{Hinz}}, \citenamefont {{Gil de Paz}}, \citenamefont
  {{Athanassoula}}, \citenamefont {{Bosma}}, \citenamefont {{Buta}},
  \citenamefont {{Cisternas}}, \citenamefont {{Ho}}, \citenamefont
  {{Holwerda}}, \citenamefont {{Skibba}}, \citenamefont {{Laurikainen}},
  \citenamefont {{Salo}}, \citenamefont {{Gadotti}}, \citenamefont {{Laine}},
  \citenamefont {{Erroz-Ferrer}}, \citenamefont {{Comer{\'o}n}}, \citenamefont
  {{Men{\'e}ndez-Delmestre}}, \citenamefont {{Seibert}},\ and\ \citenamefont
  {{Mizusawa}}}]{meidt}%
  \BibitemOpen
  \bibfield  {author} {\bibinfo {author} {\bibfnamefont {S.~E.}\ \bibnamefont
  {{Meidt}}}, \bibinfo {author} {\bibfnamefont {E.}~\bibnamefont
  {{Schinnerer}}}, \bibinfo {author} {\bibfnamefont {G.}~\bibnamefont {{van de
  Ven}}}, \bibinfo {author} {\bibfnamefont {D.}~\bibnamefont {{Zaritsky}}},
  \bibinfo {author} {\bibfnamefont {R.}~\bibnamefont {{Peletier}}}, \bibinfo
  {author} {\bibfnamefont {J.~H.}\ \bibnamefont {{Knapen}}}, \bibinfo {author}
  {\bibfnamefont {K.}~\bibnamefont {{Sheth}}}, \bibinfo {author} {\bibfnamefont
  {M.}~\bibnamefont {{Regan}}}, \bibinfo {author} {\bibfnamefont
  {M.}~\bibnamefont {{Querejeta}}}, \bibinfo {author} {\bibfnamefont {J.-C.}\
  \bibnamefont {{Mu{\~n}oz-Mateos}}}, \bibinfo {author} {\bibfnamefont
  {T.}~\bibnamefont {{Kim}}}, \bibinfo {author} {\bibfnamefont {J.~L.}\
  \bibnamefont {{Hinz}}}, \bibinfo {author} {\bibfnamefont {A.}~\bibnamefont
  {{Gil de Paz}}}, \bibinfo {author} {\bibfnamefont {E.}~\bibnamefont
  {{Athanassoula}}}, \bibinfo {author} {\bibfnamefont {A.}~\bibnamefont
  {{Bosma}}}, \bibinfo {author} {\bibfnamefont {R.~J.}\ \bibnamefont {{Buta}}},
  \bibinfo {author} {\bibfnamefont {M.}~\bibnamefont {{Cisternas}}}, \bibinfo
  {author} {\bibfnamefont {L.~C.}\ \bibnamefont {{Ho}}}, \bibinfo {author}
  {\bibfnamefont {B.}~\bibnamefont {{Holwerda}}}, \bibinfo {author}
  {\bibfnamefont {R.}~\bibnamefont {{Skibba}}}, \bibinfo {author}
  {\bibfnamefont {E.}~\bibnamefont {{Laurikainen}}}, \bibinfo {author}
  {\bibfnamefont {H.}~\bibnamefont {{Salo}}}, \bibinfo {author} {\bibfnamefont
  {D.~A.}\ \bibnamefont {{Gadotti}}}, \bibinfo {author} {\bibfnamefont
  {J.}~\bibnamefont {{Laine}}}, \bibinfo {author} {\bibfnamefont
  {S.}~\bibnamefont {{Erroz-Ferrer}}}, \bibinfo {author} {\bibfnamefont
  {S.}~\bibnamefont {{Comer{\'o}n}}}, \bibinfo {author} {\bibfnamefont
  {K.}~\bibnamefont {{Men{\'e}ndez-Delmestre}}}, \bibinfo {author}
  {\bibfnamefont {M.}~\bibnamefont {{Seibert}}}, \ and\ \bibinfo {author}
  {\bibfnamefont {T.}~\bibnamefont {{Mizusawa}}},\ }\href {\doibase
  10.1088/0004-637X/788/2/144} {\bibfield  {journal} {\bibinfo  {journal}
  {Astrophys. J.}\ }\textbf {\bibinfo {volume} {788}},\ \bibinfo {eid} {144}
  (\bibinfo {year} {2014})},\ \Eprint {http://arxiv.org/abs/1402.5210}
  {arXiv:1402.5210} \BibitemShut {NoStop}%
\bibitem [{\citenamefont {{Schombert}}\ and\ \citenamefont
  {{McGaugh}}(2014{\natexlab{b}})}]{SMpaper3}%
  \BibitemOpen
  \bibfield  {author} {\bibinfo {author} {\bibfnamefont {J.}~\bibnamefont
  {{Schombert}}}\ and\ \bibinfo {author} {\bibfnamefont {S.}~\bibnamefont
  {{McGaugh}}},\ }\href {\doibase 10.1017/pasa.2014.32} {\bibfield  {journal}
  {\bibinfo  {journal} {Pub. Astron. Soc. Australia}\ }\textbf {\bibinfo
  {volume} {31}},\ \bibinfo {eid} {e036} (\bibinfo {year}
  {2014}{\natexlab{b}})},\ \Eprint {http://arxiv.org/abs/1407.6778}
  {arXiv:1407.6778} \BibitemShut {NoStop}%
\bibitem [{\citenamefont {{McGaugh}}\ and\ \citenamefont
  {{Schombert}}(2014)}]{MS14}%
  \BibitemOpen
  \bibfield  {author} {\bibinfo {author} {\bibfnamefont {S.~S.}\ \bibnamefont
  {{McGaugh}}}\ and\ \bibinfo {author} {\bibfnamefont {J.~M.}\ \bibnamefont
  {{Schombert}}},\ }\href {\doibase 10.1088/0004-6256/148/5/77} {\bibfield
  {journal} {\bibinfo  {journal} {Astron. J.}\ }\textbf {\bibinfo {volume}
  {148}},\ \bibinfo {eid} {77} (\bibinfo {year} {2014})},\ \Eprint
  {http://arxiv.org/abs/1407.1839} {arXiv:1407.1839} \BibitemShut {NoStop}%
\bibitem [{\citenamefont {{Martinsson}}\ \emph {et~al.}(2013)\citenamefont
  {{Martinsson}}, \citenamefont {{Verheijen}}, \citenamefont {{Westfall}},
  \citenamefont {{Bershady}}, \citenamefont {{Andersen}},\ and\ \citenamefont
  {{Swaters}}}]{DiskMass7}%
  \BibitemOpen
  \bibfield  {author} {\bibinfo {author} {\bibfnamefont {T.~P.~K.}\
  \bibnamefont {{Martinsson}}}, \bibinfo {author} {\bibfnamefont {M.~A.~W.}\
  \bibnamefont {{Verheijen}}}, \bibinfo {author} {\bibfnamefont {K.~B.}\
  \bibnamefont {{Westfall}}}, \bibinfo {author} {\bibfnamefont {M.~A.}\
  \bibnamefont {{Bershady}}}, \bibinfo {author} {\bibfnamefont {D.~R.}\
  \bibnamefont {{Andersen}}}, \ and\ \bibinfo {author} {\bibfnamefont {R.~A.}\
  \bibnamefont {{Swaters}}},\ }\href {\doibase 10.1051/0004-6361/201321390}
  {\bibfield  {journal} {\bibinfo  {journal} {Astron. Astrophys.}\ }\textbf
  {\bibinfo {volume} {557}},\ \bibinfo {eid} {A131} (\bibinfo {year} {2013})},\
  \Eprint {http://arxiv.org/abs/1308.0336} {arXiv:1308.0336 [astro-ph.CO]}
  \BibitemShut {NoStop}%
\bibitem [{\citenamefont {{Draine}}(2011)}]{Draine}%
  \BibitemOpen
  \bibfield  {author} {\bibinfo {author} {\bibfnamefont {B.~T.}\ \bibnamefont
  {{Draine}}},\ }\href@noop {} {\emph {\bibinfo {title} {Physics of the
  Interstellar and Intergalactic Medium}}}\ (\bibinfo  {publisher} {Princeton,
  NJ, Princeton University Press},\ \bibinfo {year} {2011})\BibitemShut
  {NoStop}%
\bibitem [{\citenamefont {{Aver}}\ \emph {et~al.}(2010)\citenamefont {{Aver}},
  \citenamefont {{Olive}},\ and\ \citenamefont {{Skillman}}}]{BBNHelium}%
  \BibitemOpen
  \bibfield  {author} {\bibinfo {author} {\bibfnamefont {E.}~\bibnamefont
  {{Aver}}}, \bibinfo {author} {\bibfnamefont {K.~A.}\ \bibnamefont {{Olive}}},
  \ and\ \bibinfo {author} {\bibfnamefont {E.~D.}\ \bibnamefont {{Skillman}}},\
  }\href {\doibase 10.1088/1475-7516/2010/05/003} {\bibfield  {journal}
  {\bibinfo  {journal} {JCAP}\ }\textbf {\bibinfo {volume} {5}},\ \bibinfo
  {eid} {003} (\bibinfo {year} {2010})},\ \Eprint
  {http://arxiv.org/abs/1001.5218} {arXiv:1001.5218} \BibitemShut {NoStop}%
\bibitem [{\citenamefont {{Casertano}}(1983)}]{Casertano}%
  \BibitemOpen
  \bibfield  {author} {\bibinfo {author} {\bibfnamefont {S.}~\bibnamefont
  {{Casertano}}},\ }\href {\doibase 10.1093/mnras/203.3.735} {\bibfield
  {journal} {\bibinfo  {journal} {Mon. Not. Royal Astron. Soc.}\ }\textbf
  {\bibinfo {volume} {203}},\ \bibinfo {pages} {735} (\bibinfo {year}
  {1983})}\BibitemShut {NoStop}%
\bibitem [{\citenamefont {{Binney}}\ and\ \citenamefont
  {{Tremaine}}(1987)}]{BT}%
  \BibitemOpen
  \bibfield  {author} {\bibinfo {author} {\bibfnamefont {J.}~\bibnamefont
  {{Binney}}}\ and\ \bibinfo {author} {\bibfnamefont {S.}~\bibnamefont
  {{Tremaine}}},\ }\href@noop {} {\emph {\bibinfo {title} {Galactic
  Dynamics}}}\ (\bibinfo  {publisher} {Princeton, NJ, Princeton University
  Press},\ \bibinfo {year} {1987})\BibitemShut {NoStop}%
\bibitem [{\citenamefont {{van der Hulst}}\ \emph {et~al.}(1992)\citenamefont
  {{van der Hulst}}, \citenamefont {{Terlouw}}, \citenamefont {{Begeman}},
  \citenamefont {{Zwitser}},\ and\ \citenamefont {{Roelfsema}}}]{GIPSY}%
  \BibitemOpen
  \bibfield  {author} {\bibinfo {author} {\bibfnamefont {J.~M.}\ \bibnamefont
  {{van der Hulst}}}, \bibinfo {author} {\bibfnamefont {J.~P.}\ \bibnamefont
  {{Terlouw}}}, \bibinfo {author} {\bibfnamefont {K.~G.}\ \bibnamefont
  {{Begeman}}}, \bibinfo {author} {\bibfnamefont {W.}~\bibnamefont
  {{Zwitser}}}, \ and\ \bibinfo {author} {\bibfnamefont {P.~R.}\ \bibnamefont
  {{Roelfsema}}},\ }in\ \href@noop {} {\emph {\bibinfo {booktitle}
  {Astronomical Data Analysis Software and Systems I}}},\ \bibinfo {series}
  {Astronomical Society of the Pacific Conference Series}, Vol.~\bibinfo
  {volume} {25},\ \bibinfo {editor} {edited by\ \bibinfo {editor}
  {\bibfnamefont {D.~M.}\ \bibnamefont {{Worrall}}}, \bibinfo {editor}
  {\bibfnamefont {C.}~\bibnamefont {{Biemesderfer}}}, \ and\ \bibinfo {editor}
  {\bibfnamefont {J.}~\bibnamefont {{Barnes}}}}\ (\bibinfo {year} {1992})\ p.\
  \bibinfo {pages} {131}\BibitemShut {NoStop}%
\bibitem [{\citenamefont {{Bell}}\ and\ \citenamefont {{de Jong}}(2001)}]{BdJ}%
  \BibitemOpen
  \bibfield  {author} {\bibinfo {author} {\bibfnamefont {E.~F.}\ \bibnamefont
  {{Bell}}}\ and\ \bibinfo {author} {\bibfnamefont {R.~S.}\ \bibnamefont {{de
  Jong}}},\ }\href {\doibase 10.1086/319728} {\bibfield  {journal} {\bibinfo
  {journal} {Astrophys. J.}\ }\textbf {\bibinfo {volume} {550}},\ \bibinfo
  {pages} {212} (\bibinfo {year} {2001})}\BibitemShut {NoStop}%
\bibitem [{\citenamefont {{Lelli}}\ \emph
  {et~al.}(2016{\natexlab{c}})\citenamefont {{Lelli}}, \citenamefont
  {{McGaugh}},\ and\ \citenamefont {{Schombert}}}]{SPARCII}%
  \BibitemOpen
  \bibfield  {author} {\bibinfo {author} {\bibfnamefont {F.}~\bibnamefont
  {{Lelli}}}, \bibinfo {author} {\bibfnamefont {S.~S.}\ \bibnamefont
  {{McGaugh}}}, \ and\ \bibinfo {author} {\bibfnamefont {J.~M.}\ \bibnamefont
  {{Schombert}}},\ }\href {\doibase unknown} {\bibfield  {journal} {\bibinfo
  {journal} {Astrophys. J.}\ ,\ \bibinfo {pages} {in prep.}} (\bibinfo {year}
  {2016}{\natexlab{c}})}\BibitemShut {NoStop}%
\bibitem [{\citenamefont {{van Albada}}\ and\ \citenamefont
  {{Sancisi}}(1986)}]{vAS1986}%
  \BibitemOpen
  \bibfield  {author} {\bibinfo {author} {\bibfnamefont {T.~S.}\ \bibnamefont
  {{van Albada}}}\ and\ \bibinfo {author} {\bibfnamefont {R.}~\bibnamefont
  {{Sancisi}}},\ }\href {\doibase 10.1098/rsta.1986.0128} {\bibfield  {journal}
  {\bibinfo  {journal} {Philosophical Transactions of the Royal Society of
  London Series A}\ }\textbf {\bibinfo {volume} {320}},\ \bibinfo {pages} {447}
  (\bibinfo {year} {1986})}\BibitemShut {NoStop}%
\bibitem [{\citenamefont {{de Blok}}\ and\ \citenamefont
  {{McGaugh}}(1997)}]{dBM97}%
  \BibitemOpen
  \bibfield  {author} {\bibinfo {author} {\bibfnamefont {W.~J.~G.}\
  \bibnamefont {{de Blok}}}\ and\ \bibinfo {author} {\bibfnamefont {S.~S.}\
  \bibnamefont {{McGaugh}}},\ }\href {\doibase 10.1093/mnras/290.3.533}
  {\bibfield  {journal} {\bibinfo  {journal} {Mon. Not. Royal Astron. Soc.}\
  }\textbf {\bibinfo {volume} {290}},\ \bibinfo {pages} {533} (\bibinfo {year}
  {1997})},\ \Eprint {http://arxiv.org/abs/astro-ph/9704274} {astro-ph/9704274}
  \BibitemShut {NoStop}%
\bibitem [{\citenamefont {{de Blok}}\ and\ \citenamefont
  {{McGaugh}}(1996)}]{dBM96}%
  \BibitemOpen
  \bibfield  {author} {\bibinfo {author} {\bibfnamefont {W.~J.~G.}\
  \bibnamefont {{de Blok}}}\ and\ \bibinfo {author} {\bibfnamefont {S.~S.}\
  \bibnamefont {{McGaugh}}},\ }\href {\doibase 10.1086/310266} {\bibfield
  {journal} {\bibinfo  {journal} {Astrophys. J.l}\ }\textbf {\bibinfo {volume}
  {469}},\ \bibinfo {pages} {L89} (\bibinfo {year} {1996})},\ \Eprint
  {http://arxiv.org/abs/astro-ph/9607042} {astro-ph/9607042} \BibitemShut
  {NoStop}%
\bibitem [{\citenamefont {{Lelli}}\ \emph {et~al.}(2013)\citenamefont
  {{Lelli}}, \citenamefont {{Fraternali}},\ and\ \citenamefont
  {{Verheijen}}}]{RCgradient}%
  \BibitemOpen
  \bibfield  {author} {\bibinfo {author} {\bibfnamefont {F.}~\bibnamefont
  {{Lelli}}}, \bibinfo {author} {\bibfnamefont {F.}~\bibnamefont
  {{Fraternali}}}, \ and\ \bibinfo {author} {\bibfnamefont {M.}~\bibnamefont
  {{Verheijen}}},\ }\href {\doibase 10.1093/mnrasl/slt053} {\bibfield
  {journal} {\bibinfo  {journal} {Mon. Not. Royal Astron. Soc.}\ }\textbf
  {\bibinfo {volume} {433}},\ \bibinfo {pages} {L30} (\bibinfo {year}
  {2013})},\ \Eprint {http://arxiv.org/abs/1304.4250} {arXiv:1304.4250}
  \BibitemShut {NoStop}%
\bibitem [{\citenamefont {{Lelli}}\ \emph
  {et~al.}(2016{\natexlab{d}})\citenamefont {{Lelli}}, \citenamefont
  {{McGaugh}}, \citenamefont {{Schombert}},\ and\ \citenamefont
  {{Pawlowski}}}]{CentralDensRelation}%
  \BibitemOpen
  \bibfield  {author} {\bibinfo {author} {\bibfnamefont {F.}~\bibnamefont
  {{Lelli}}}, \bibinfo {author} {\bibfnamefont {S.~S.}\ \bibnamefont
  {{McGaugh}}}, \bibinfo {author} {\bibfnamefont {J.~M.}\ \bibnamefont
  {{Schombert}}}, \ and\ \bibinfo {author} {\bibfnamefont {M.~S.}\ \bibnamefont
  {{Pawlowski}}},\ }\href {\doibase 10.3847/2041-8205/827/1/L19} {\bibfield
  {journal} {\bibinfo  {journal} {Astrophys. J.}\ }\textbf {\bibinfo {volume}
  {827}},\ \bibinfo {eid} {L19} (\bibinfo {year} {2016}{\natexlab{d}})},\
  \Eprint {http://arxiv.org/abs/1607.02145} {arXiv:1607.02145} \BibitemShut
  {NoStop}%
\bibitem [{\citenamefont {{McGaugh}}(2008)}]{M08}%
  \BibitemOpen
  \bibfield  {author} {\bibinfo {author} {\bibfnamefont {S.~S.}\ \bibnamefont
  {{McGaugh}}},\ }\href {\doibase 10.1086/589148} {\bibfield  {journal}
  {\bibinfo  {journal} {Astrophys. J.}\ }\textbf {\bibinfo {volume} {683}},\
  \bibinfo {pages} {137} (\bibinfo {year} {2008})},\ \Eprint
  {http://arxiv.org/abs/0804.1314} {arXiv:0804.1314} \BibitemShut {NoStop}%
\bibitem [{\citenamefont {{van den Bosch}}\ and\ \citenamefont
  {{Swaters}}(2001)}]{vdBSwat}%
  \BibitemOpen
  \bibfield  {author} {\bibinfo {author} {\bibfnamefont {F.~C.}\ \bibnamefont
  {{van den Bosch}}}\ and\ \bibinfo {author} {\bibfnamefont {R.~A.}\
  \bibnamefont {{Swaters}}},\ }\href {\doibase
  10.1046/j.1365-8711.2001.04456.x} {\bibfield  {journal} {\bibinfo  {journal}
  {Mon. Not. Royal Astron. Soc.}\ }\textbf {\bibinfo {volume} {325}},\ \bibinfo
  {pages} {1017} (\bibinfo {year} {2001})},\ \Eprint
  {http://arxiv.org/abs/astro-ph/0006048} {astro-ph/0006048} \BibitemShut
  {NoStop}%
\bibitem [{\citenamefont {{Oman}}\ \emph {et~al.}(2016)\citenamefont {{Oman}},
  \citenamefont {{Navarro}}, \citenamefont {{Sales}}, \citenamefont
  {{Fattahi}}, \citenamefont {{Frenk}}, \citenamefont {{Sawala}}, \citenamefont
  {{Schaller}},\ and\ \citenamefont {{White}}}]{OmanApostle}%
  \BibitemOpen
  \bibfield  {author} {\bibinfo {author} {\bibfnamefont {K.~A.}\ \bibnamefont
  {{Oman}}}, \bibinfo {author} {\bibfnamefont {J.~F.}\ \bibnamefont
  {{Navarro}}}, \bibinfo {author} {\bibfnamefont {L.~V.}\ \bibnamefont
  {{Sales}}}, \bibinfo {author} {\bibfnamefont {A.}~\bibnamefont {{Fattahi}}},
  \bibinfo {author} {\bibfnamefont {C.~S.}\ \bibnamefont {{Frenk}}}, \bibinfo
  {author} {\bibfnamefont {T.}~\bibnamefont {{Sawala}}}, \bibinfo {author}
  {\bibfnamefont {M.}~\bibnamefont {{Schaller}}}, \ and\ \bibinfo {author}
  {\bibfnamefont {S.~D.~M.}\ \bibnamefont {{White}}},\ }\href {\doibase
  10.1093/mnras/stw1251} {\bibfield  {journal} {\bibinfo  {journal} {Mon. Not.
  Royal Astron. Soc.}\ }\textbf {\bibinfo {volume} {460}},\ \bibinfo {pages}
  {3610} (\bibinfo {year} {2016})},\ \Eprint {http://arxiv.org/abs/1601.01026}
  {arXiv:1601.01026} \BibitemShut {NoStop}%
\bibitem [{\citenamefont {{Pineda}}\ \emph {et~al.}(2016)\citenamefont
  {{Pineda}}, \citenamefont {{Hayward}}, \citenamefont {{Springel}},\ and\
  \citenamefont {{Mendes de Oliveira}}}]{Pinhead}%
  \BibitemOpen
  \bibfield  {author} {\bibinfo {author} {\bibfnamefont {J.~C.~B.}\
  \bibnamefont {{Pineda}}}, \bibinfo {author} {\bibfnamefont {C.~C.}\
  \bibnamefont {{Hayward}}}, \bibinfo {author} {\bibfnamefont {V.}~\bibnamefont
  {{Springel}}}, \ and\ \bibinfo {author} {\bibfnamefont {C.}~\bibnamefont
  {{Mendes de Oliveira}}},\ }\href@noop {} {\bibfield  {journal} {\bibinfo
  {journal} {ArXiv e-prints}\ } (\bibinfo {year} {2016})},\ \Eprint
  {http://arxiv.org/abs/1602.07690} {arXiv:1602.07690} \BibitemShut {NoStop}%
\bibitem [{\citenamefont {{de Blok}}(2010)}]{dBcorecusp}%
  \BibitemOpen
  \bibfield  {author} {\bibinfo {author} {\bibfnamefont {W.~J.~G.}\
  \bibnamefont {{de Blok}}},\ }\href {\doibase 10.1155/2010/789293} {\bibfield
  {journal} {\bibinfo  {journal} {Advances in Astronomy}\ }\textbf {\bibinfo
  {volume} {2010}},\ \bibinfo {eid} {789293} (\bibinfo {year} {2010})},\
  \Eprint {http://arxiv.org/abs/0910.3538} {arXiv:0910.3538} \BibitemShut
  {NoStop}%
\bibitem [{\citenamefont {{Navarro}}\ \emph {et~al.}(1997)\citenamefont
  {{Navarro}}, \citenamefont {{Frenk}},\ and\ \citenamefont {{White}}}]{NFW}%
  \BibitemOpen
  \bibfield  {author} {\bibinfo {author} {\bibfnamefont {J.~F.}\ \bibnamefont
  {{Navarro}}}, \bibinfo {author} {\bibfnamefont {C.~S.}\ \bibnamefont
  {{Frenk}}}, \ and\ \bibinfo {author} {\bibfnamefont {S.~D.~M.}\ \bibnamefont
  {{White}}},\ }\href {\doibase 10.1086/304888} {\bibfield  {journal} {\bibinfo
   {journal} {Astrophys. J.}\ }\textbf {\bibinfo {volume} {490}},\ \bibinfo
  {pages} {493} (\bibinfo {year} {1997})}\BibitemShut {NoStop}%
\bibitem [{\citenamefont {{McGaugh}}(2015)}]{McG2014}%
  \BibitemOpen
  \bibfield  {author} {\bibinfo {author} {\bibfnamefont {S.~S.}\ \bibnamefont
  {{McGaugh}}},\ }\href {\doibase 10.1139/cjp-2014-0203} {\bibfield  {journal}
  {\bibinfo  {journal} {Canadian Journal of Physics}\ }\textbf {\bibinfo
  {volume} {93}},\ \bibinfo {pages} {250} (\bibinfo {year} {2015})},\ \Eprint
  {http://arxiv.org/abs/1404.7525} {arXiv:1404.7525} \BibitemShut {NoStop}%
\bibitem [{\citenamefont {{Di Cintio}}\ and\ \citenamefont
  {{Lelli}}(2016)}]{DCL}%
  \BibitemOpen
  \bibfield  {author} {\bibinfo {author} {\bibfnamefont {A.}~\bibnamefont {{Di
  Cintio}}}\ and\ \bibinfo {author} {\bibfnamefont {F.}~\bibnamefont
  {{Lelli}}},\ }\href {\doibase 10.1093/mnrasl/slv185} {\bibfield  {journal}
  {\bibinfo  {journal} {Mon.Not. Royal Astron. Soc.}\ }\textbf {\bibinfo
  {volume} {456}},\ \bibinfo {pages} {L127} (\bibinfo {year} {2016})},\ \Eprint
  {http://arxiv.org/abs/1511.06616} {arXiv:1511.06616} \BibitemShut {NoStop}%
\bibitem [{\citenamefont {{Desmond}}(2016)}]{Desmond}%
  \BibitemOpen
  \bibfield  {author} {\bibinfo {author} {\bibfnamefont {H.}~\bibnamefont
  {{Desmond}}},\ }\href@noop {} {\bibfield  {journal} {\bibinfo  {journal}
  {arXiv:1607.01800}\ } (\bibinfo {year} {2016})},\ \Eprint
  {http://arxiv.org/abs/1607.01800} {arXiv:1607.01800} \BibitemShut {NoStop}%
\bibitem [{\citenamefont {{Zhao}}\ and\ \citenamefont
  {{Li}}(2010)}]{darkfluid}%
  \BibitemOpen
  \bibfield  {author} {\bibinfo {author} {\bibfnamefont {H.}~\bibnamefont
  {{Zhao}}}\ and\ \bibinfo {author} {\bibfnamefont {B.}~\bibnamefont {{Li}}},\
  }\href {\doibase 10.1088/0004-637X/712/1/130} {\bibfield  {journal} {\bibinfo
   {journal} {Astrophys. J.}\ }\textbf {\bibinfo {volume} {712}},\ \bibinfo
  {pages} {130} (\bibinfo {year} {2010})},\ \Eprint
  {http://arxiv.org/abs/0804.1588} {arXiv:0804.1588} \BibitemShut {NoStop}%
\bibitem [{\citenamefont {{Khoury}}(2015)}]{Khoury2015}%
  \BibitemOpen
  \bibfield  {author} {\bibinfo {author} {\bibfnamefont {J.}~\bibnamefont
  {{Khoury}}},\ }\href {\doibase 10.1103/PhysRevD.91.024022} {\bibfield
  {journal} {\bibinfo  {journal} {Phys. Rev. D}\ }\textbf {\bibinfo {volume}
  {91}},\ \bibinfo {eid} {024022} (\bibinfo {year} {2015})},\ \Eprint
  {http://arxiv.org/abs/1409.0012} {arXiv:1409.0012 [hep-th]} \BibitemShut
  {NoStop}%
\bibitem [{\citenamefont {{Blanchet}}(2007)}]{blanchet}%
  \BibitemOpen
  \bibfield  {author} {\bibinfo {author} {\bibfnamefont {L.}~\bibnamefont
  {{Blanchet}}},\ }\href {\doibase 10.1088/0264-9381/24/14/001} {\bibfield
  {journal} {\bibinfo  {journal} {Classical and Quantum Gravity}\ }\textbf
  {\bibinfo {volume} {24}},\ \bibinfo {pages} {3529} (\bibinfo {year}
  {2007})}\BibitemShut {NoStop}%
\bibitem [{\citenamefont {{Milgrom}}(1983)}]{milgrom83}%
  \BibitemOpen
  \bibfield  {author} {\bibinfo {author} {\bibfnamefont {M.}~\bibnamefont
  {{Milgrom}}},\ }\href@noop {} {\bibfield  {journal} {\bibinfo  {journal}
  {{Astrophys. J.}}\ }\textbf {\bibinfo {volume} {270}},\ \bibinfo {pages}
  {371} (\bibinfo {year} {1983})}\BibitemShut {NoStop}%
\end{thebibliography}

%

\end{document}